\documentclass[conference]{IEEEtran}
\IEEEoverridecommandlockouts
\usepackage{cite}
\usepackage{amsmath,amssymb,amsfonts}
\usepackage{algorithmic}
\usepackage{graphicx}
\usepackage{textcomp}
\usepackage{xcolor}

\def\BibTeX{{\rm B\kern-.05em{\sc i\kern-.025em b}\kern-.08em
    T\kern-.1667em\lower.7ex\hbox{E}\kern-.125emX}}

\newcommand{\RNum}[1]{\uppercase\expandafter{\romannumeral #1\relax}}
    
\begin{document}

\title{Ratings of European and South American Football
Leagues Based on Glicko-2 with Modifications}

\author{\IEEEauthorblockN{1\textsuperscript{st} Andrei Shelopugin}
\IEEEauthorblockA{
\textit{Independent Researcher}\\
shelopuginandrey@gmail.com}
\and
\IEEEauthorblockN{2\textsuperscript{nd} Alexander Sirotkin}
\IEEEauthorblockA{
\textit{HSE University}\\
Saint Petersburg, Russia \\
alexander.sirotkin@gmail.com}

}

\maketitle

\thispagestyle{empty}
\begin{abstract}
One of the key problems in the field of soccer analytics is predicting how a player performance changes when transitioning from one league to another. One potential solution to address this issue lies in the evaluation of the respective league strength.

This article endeavors to compute club ratings of the first and second European and South American leagues. In order to calculate these ratings, the authors have designed the Glicko-2 rating system-based approach, which overcomes some Glicko-2 limitations. Particularly, the authors took into consideration the probability of the draw, the home-field advantage, and the property of teams to become stronger or weaker following their league transitions. Furthermore, authors have constructed a predictive model for forecasting match results based on the number of goals scored in previous matches. The metrics utilized in the analysis reveal that the Glicko-2 based approach exhibits a marginally superior level of accuracy when compared to the commonly used Poisson regression-based approach. In addition, Glicko-2 based ratings offer greater interpretability and can find application in various analytics tasks, such as predicting soccer player metrics for forthcoming seasons or the detailed analysis of a player performance in preceding matches.

The implementation of the approach is available on github.com/andreyshelopugin/GlickoSoccer.

\end{abstract}


\section{Introduction}

\subsection{Player Selection Problem}
One of the main challenges faced by football managers, scouts, and analysts is the assessment of players’ abilities. This problem can be addressed through video analysis of players game or by analysis of players performance metrics. In the latter scenario, it becomes imperative for an analyst to understand the resistance level of different championships for optimal decision-making process for players hiring.

\subsection{Problem of Matches Outcomes Prediction}
It is necessary to create a ranking model for predicting matches outcomes. This ranking model affords us the capability to simulate forthcoming match outcomes, thereby assisting football managers in strategic transfer planning. Moreover, our ranking model can also be used to impose constraints on a draw simulator for final tournaments, enhancing group parity, and subsequently elevating spectator engagement.

\subsection{Rating UEFA Problem}
The UEFA rating is the most acknowledged system in soccer, therefore we consider its weak spots. Firstly, it neglects to consider the strength of opponents, treating victories over favorites and weaker teams equally. Additionally, it places significant emphasis on past performance, which can impact a team’s current points and the results of a draw round. In the case of the performance drop, strong opponents would be selected for a weakened team, hindering points accumulation for the latter. Essentially, the UEFA rating system employs the same weight coefficients for matches, irrespective of the season or year, thus failing to accurately reflect teams’ current strengths.

In addition to the aforementioned issues, there exist several less significant drawbacks associated with the UEFA rating system. For instance, the rating points lack a robust mathematical foundation, rendering the UEFA rating challenging to interpret.

This rating system lacks the capability to effectively assess the strength of a national league. The UEFA rating of a country is determined solely by the performance of its strongest teams. Therefore, we have no information concerning the performance of weaker teams in the first league, as well as the second and lower divisions.

\section{Matches outcomes prediction based on gradient boosting}

\subsection{Matches Outcomes Prediction Idea}
Initially we have designed a model for predicting matches outcomes. Our approach leverages a widely recognized concept for forecasting football match results, involving the prediction of the number of goals each team will score against their respective opponents. Subsequently, we used the assumption that the goals in football conform to a Poisson distribution. This assumption enables the utilization of the probability mass function associated with a Skellam distribution:

\begin{equation}
    p(k, \mu_1, \mu_2) = exp(-(\mu_1 + \mu_2)) (\frac{\mu_1}{\mu_2})^{\frac{k}{2}}I_k(2\sqrt{\mu_1 \mu_2})
\end{equation}

In the formula $I_k$ is the modified Bessel function, $k$ is a score difference in football match. So, if $k > 0$, the value of the probability mass function equals the probability of the first team winning, for $k = 0$ we received the draw probability, and finally if $k < 0$, we get the probability of the second team winning. Parameters $\mu_1$ and $\mu_2$ are the means of two Poisson-distributed random variables. In our case we use goals predictions as $\mu_1$ and $\mu_2$.

\subsection{Implementation of gradient boosting based model}
Gradient boosting models have demonstrated impressive performance when applied to tabular data. Consequently, we have constructed a machine learning model utilizing the LightGBM framework to predict the goal count for both competing teams. We also experimented with the CatBoost model; however, it yielded inferior results. The goals in football conform to the Poisson distribution, therefore we set LightGBM parameter "objective" to "poisson".

The algorithm was trained using the following features, with each row in the dataset corresponding to a participant in a match:

- a team plays at home (true/false);

- a match happens during pandemic (true/false);

- the average number of goals in the last 5, 10, 20, 30
matches (numeric);
- the average number of team opponent missed goals in the
last 5, 10, 20, 30 matches (numeric);

- a tournament, e.g. Italian Serie A, Champions League,
Coppa Italia (categorical);

- a league of team/opponent, e.g. Italian Serie A (categorical);

- the average/median number of goals scored at home (if a
team plays at home in the current match) or away (if a team plays away in the current match) in the last 5, 10, 20, 30 matches (numeric);

- the average/median number of goals missed by the opponent team when playing away(if a team plays at home in the current match) or at home field (if a team plays away in the current match) in the last 5, 10, 20, 30 matches (numeric).

\subsection{Data}
The data was collected from flashscore.com. The data consists of the matches of Europe and South America first and second leagues, matches of national cups, and international tournaments such as Champions League or Copa Libertadores. It’s important to note that matches involving teams from lower-ranking leagues, specifically those in the third division and below, have been deliberately excluded. This includes instances such as national cup matches featuring third-tier teams. Additionally, matches decided by forfeit victories have also been omitted from the dataset. The dataset contains approximately 366 000 matches played between 2010/2011 and 2022/2023 seasons.

\subsection{Train Test Split}
The model was trained using match data from the 2010/2011 through the 2020/2021 seasons, while matches that took place during the 2021/2022 and 2022/2023 seasons were set aside and selected for use as a test set.

\section{Matches outcomes prediction based on Glicko-2 with modifications}

\subsection{Rating Systems}
Today there are a lot of rating systems designed for predicting matches outcomes: Elo~\cite{elo1967proposed}, Glicko~\cite{glickman2012example}, TrueSkill~\cite{herbrich2006trueskill}. Nevertheless, it is necessary that opponents are placed on a level playing field, otherwise, these models tend to exhibit suboptimal performance. This inherent limitation makes these models inapplicable for effectively forecasting outcomes in team sports like football, basketball, or ice hockey. Notably, existing rating systems neglect crucial factors such as home team advantage, player injuries, player transfers, and more.

In this article we propose the model for evaluating the strength of the European football Leagues. Although some models calculate ratings of football Leagues, such as Club Elo~\cite{clubelo} and UEFA rating~\cite{uefa}, existing approaches suffer serious drawbacks. 

\subsection{Glicko-2}
To address the aforementioned challenges, we propose a model that leverages the Glicko-2 algorithm, specifically tailored for football applications.

The Glicko-2 model has various applications across various domains in both sports and computer science, including designing rating systems for chess~\cite{vevcek2014chess} and online games~\cite{dehpanah2021evaluation}, volleyball~\cite{glickman2018comparison}, programming competitions~\cite{ebtekar2021elo}, obstacle courses races and other multiplayer competitions~\cite{ebtekar2021elo}. 
Besides sports, Glicko-2 is widely used for many other purposes, for example, online courses evaluation~\cite{park2021online}, the evaluation of the discriminator and generator performance in generative models~\cite{olsson2018skill} and the comparison of different machine learning benchmarks~\cite{cardoso2020decoding}.

Glicko-2 represents the strength of each team as a prob- ability distribution. Initially, each team is associated with a normal prior distribution. As match outcomes accumulate, teams are approximated to a posterior distribution. When the team wins its rating increases. This increase is influenced by two key factors: firstly, the difference in ratings, resulting in a more substantial increase when a poor-rated team wins, and secondly, rating deviations, which lead to rapid ratings adjustments when there is limited historical match data available for the opponents.

\subsection{Draw Probability}
The initial challenge pertains to the potential occurrence of a draw in football matches. The mathematical expectation formulas utilized in Glicko-2 do not facilitate the assessment of draw probabilities. The draw probability is influenced not solely by the disparity in skill levels between competing teams but also by the scoring tendencies of the teams involved. For instance, the probability of a draw would be marginally higher in a match featuring two teams with a history of low-scoring performances.

The original Glicko-2 version calculates win probability in this way:

\begin{equation} E(\mu, \mu_j, \phi_j) = \frac {1}{1 + exp(-g(\phi_j)(\mu - \mu_j))} 
\end{equation}

We have modified it in the following way:

\begin{equation} 
E(\mu, \mu_j, \phi_j, d, s) = \frac {exp(g(\phi_j)(\mu - \mu_j))}{1 + exp(g(\phi_j)(\mu - \mu_j)) + exp(d+s)}
\end{equation}

We have added two parameters into the original formula: $s$ represents the draw probability derived from the LightGBM based model and  $d$ serves as a hyperparameter responsible for adjusting draw outcomes.

It is necessary to clarify the introduced term adjusting draw outcomes. We have used the assumption that the goals in football conform to a Poisson distribution. However the assumption is not entirely precise. A significant issue arises from the underestimation of the frequency of low-scoring matches~\cite{dixon1997modelling}, resulting in a bias in our draw predictions. Consequently, we have incorporated a hyperparameter denoted as $d$ in (3) to address this issue.

\subsection{Home Team Advantage}

Another aspect to take into account is the advantage of the home team, which is influenced by the support of their fans and the logistical challenges faced by the visiting team. Consequently, due to the diminished impact of the home team advantage during the pandemic, we conducted separate training for the home team constant for the period from March 2020 to June 2021.

The original Glicko-2 version updates rating in this way:

\begin{equation}
    \mu^{'} = \mu + \phi^{'2}g(\phi_j)(s_j - E(\mu, \mu_j, \phi_j))
\end{equation}

We have modified it for a home team (with modifications from (3)): 

\begin{equation}
\mu^{'} = \mu + \phi^{'2}g(\phi_j)(s_j - E(\mu + h, \mu_j, \phi_j, d, s)) 
\end{equation}

The parameter $h$ will be replaced by $h_{p}$ during the pandemic.
The parameters $h$ and $h_{p}$ are positive numbers representing the home field advantage. Also we have imposed the following restriction: 

\begin{equation}
    h > h_{p}
\end{equation}

\subsection{Rating Initialization}
The Glicko-2 model takes into consideration only the difference of the team ratings. To enhance its accuracy, we can introduce a novel parameter known as the "initial league rating," which characterizes the average rating within a league. This parameter does not influence the National Championship match outcomes prediction but proves useful for comparing the opponents from diverse national leagues, particularly in international Cups. It effectively prevents underfitting, ensuring more robust comparisons.

However, this approach can inadvertently lead to rating inflation. The initial league rating represents roughly the average league rating. As we mentioned earlier, our dataset includes data starting from the year 2010. Now, let’s imagine we have a team whose earliest match in our dataset dates back to the year 2015 in the second league. This implies that the team played in weaker leagues before 2015, which means its rating should be below the average rating of the second league.

To address this issue, we propose a solution: penalizing the initial league rating for teams coming from lower-ranked leagues, thereby setting them below the average league rating.
As a result, we update team ratings before each season, and the rating initialization process unfolds as follows:

\begin{equation}
r(\mu,\phi,\sigma) =
 \begin{cases}
   r(\mu_{init} + \mu_{new},\phi,\sigma) &\text{if team was promoted,}\\
   r(\mu_{init},\phi,\sigma) &\text{otherwise.}
 \end{cases}
\end{equation}

The parameter $\mu_{new}$ should be negative.

\subsection{Mid-Season}
The fact that the team rosters can change during the transfer window cannot be overlooked. To address this issue, we lever- age the rating deviation parameter from the Glicko-2 model. At the beginning of a tournament, ratings are notably influenced by every win or loss, but as more data accumulates, these ratings undergo fewer deviations. We update this parameter at the beginning of every season.

Furthermore, we introduce a parameter that accounts for the propensity of teams to either strengthen or weaken during transitions between leagues. For instance, when a team is promoted to the Premier League or relegated to the Second League, its rating experiences slight adjustments. This parameter enables us to capture the tendency of different national leagues to launch an active transfer campaign after a league change.

Consequently, we update team ratings prior to the commencement of every season using the following formula:

\begin{equation}
r(\mu,\phi,\sigma) =
 \begin{cases}
   r(\mu + \mu_{l},\phi + \phi_{s},\sigma) &\text{if team changed league,}\\
   r(\mu,\phi + \phi_{s},\sigma) &\text{otherwise.}
 \end{cases}
\end{equation}

The parameter $\phi_{s}$ should be positive to introduce a degree of uncertainty. The parameter $\mu_{l}$ should be positive for promoted teams and negative for relegated because usually the club budget increases after promotion and decreases after relegation.

The other factor that can lead to rating inflation is the fact
that the number of teams changes because of changes in league
rules etc. Thus, we normalize all ratings after each season to
prevent the inflation. Hence, the average rating of all teams
should be constant.

\subsection{Training Model}
It’s essential to highlight that we conduct training for all the parameters described across the entire league. As a result, each league possesses a distinct set of parameters, which includes $\mu_{init}$, $\mu_{new}$, $\mu_{l}$, $\phi_{s}$, $h$, $h_{p}$. 
Furthermore, we need to optimize the initial values $\phi$, $\sigma$, and $d$ from (3).

These parameters are trained through the minimization of the log loss function associated with match outcomes, enabling us to compute predictions.

While it would be theoretically desirable to assign a unique set of parameters to each team, considering factors such as stadium capacity, distance between opponent cities, and other complex variables that influence the actual home-field advantage, this approach tends to result in model overfitting.

\subsection{Results}
We have opted for the log loss metric as the criterion for assessing the quality of outcome predictions. We computed the loss functions on the test set, consisting of 60,091 matches. Below, you’ll find a comparison of the models’ performance.

\begin{table}[!ht]
\centering
\caption{Model comparison}
\label{tab:comparison}
\begin{tabular}{|l|c|}
\hline
Method & Log Loss \\ \hline
Catboost based & 0.5931   \\ 
LightGBM based & 0.5896   \\ 
Original Glicko-2 & 0.5949 \\ 
Glicko-2 with modifications & 0.5832  \\ 
\hline
\end{tabular}

\end{table}

The calculated ratings are presented in the appendix. Results are relevant for the summer of 2023.

The implementation of approach is available on github.com/andreyshelopugin/GlickoSoccer.

\subsection{Interpretation of ratings}
Club ratings offer a high level of interpretability. The table \RNum{2} provides estimations of the probabilities of a win, draw and loss based on the percentage difference in team ratings.

\begin{table}[!ht]
\centering
\caption{Interpretation of Ratings}
\label{tab:ratings}
\begin{tabular}{|r|c|c|c|}
\hline
 Ratings difference & Win & Draw & Loss \\ 
\hline
0     & 35.7 & 28.6 & 35.7 \\
10    & 37.1 & 28.5 & 34.5 \\
20    & 38.5 & 28.2 & 33.3 \\
50    & 42.5 & 27.8 & 29.7 \\
100   & 49.9 & 25.7 & 24.3 \\
200   & 63.8 & 20.5 & 15.7 \\
500   & 90.5 & 6.0  & 3.5  \\
800   & 98.1 & 1.2  & 0.7  \\
\hline
\end{tabular}
\end{table}

It’s important to note that these probabilities are approximate.  We took average parameters $\phi_j$ and $s$ from equation (3). These parameters can vary from one league to another.

\section{Example of Application League Ratings in the Task of Players Metrics Prediction}
Consider, for instance, our objective to predict various player metrics for the upcoming season, such as the number of goals a specific player will score. To make accurate predictions, we can take into account player age, position, previous performance, number of goals in the previous season and other relevant features. However, a significant challenge arises when a player undergoes transfers, moving to a different club or league during their career. How can we effectively inform the model about such transitions?

One possible solution involves introducing a categorical feature, such as "player transfers from Denmark Superliga to France Ligue 1". Nonetheless, this approach encounters the problem of dimensionality – there are relatively few transfers from the Denmark Superliga to France Ligue 1, making it challenging for the model to learn effectively. A more viable approach, and one that aligns with our method, is to generate numeric features. For example, we can create a feature that quantifies the difference in ratings between the Denmark Superliga and France Ligue 1. This approach enables the model to account for league transitions more effectively.

Additionally, we can enhance the precision of player performance analysis, particularly within the context of a single team. For instance, we can calculate the average rating of the opposing teams faced by each player. This fine-grained analysis helps differentiate player performances, as several players may exhibit similar statistics, but the strength of the opponent teams they encounter could vary significantly.

\section{Problems}

The primary limitation of the proposed approach stems from its sole reliance on match outcomes for calculations. Consequently, this approach assumes that the strength of teams remains relatively stable throughout the season. However, second league teams frequently experience significant roster changes as the season progresses.

The other challenge is the poor international matches history of teams outside the top 40 ranking UEFA. This constraint may lead to an underfitting issue, as the model lacks sufficient data for these teams. Furthermore, it’s not feasible to compare clubs from Europe with those from South America due to the scarcity of match history between teams from these two regions.

\section{Conclusion}
Despite the constraints mentioned earlier, we have developed a cost-effective model that relies only on publicly accessible data. Both methods exhibit comparable log losses; nevertheless, the Glicko-2 based model offers the advantage of utilizing the computed ratings in a wide range of analytical tasks.

% \nocite{*} % to test all bib entrys
\bibliographystyle{IEEEtran}
\bibliography{Ratings}

\begin{thebibliography}{10}
\providecommand{\url}[1]{#1}
\csname url@samestyle\endcsname
\providecommand{\newblock}{\relax}
\providecommand{\bibinfo}[2]{#2}
\providecommand{\BIBentrySTDinterwordspacing}{\spaceskip=0pt\relax}
\providecommand{\BIBentryALTinterwordstretchfactor}{4}
\providecommand{\BIBentryALTinterwordspacing}{\spaceskip=\fontdimen2\font plus
\BIBentryALTinterwordstretchfactor\fontdimen3\font minus \fontdimen4\font\relax}
\providecommand{\BIBforeignlanguage}[2]{{%
\expandafter\ifx\csname l@#1\endcsname\relax
\typeout{** WARNING: IEEEtran.bst: No hyphenation pattern has been}%
\typeout{** loaded for the language `#1'. Using the pattern for}%
\typeout{** the default language instead.}%
\else
\language=\csname l@#1\endcsname
\fi
#2}}
\providecommand{\BIBdecl}{\relax}
\BIBdecl

\bibitem{elo1967proposed}
A.~E. Elo, ``The proposed uscf rating system, its development, theory, and applications,'' \emph{Chess Life}, vol.~22, no.~8, pp. 242--247, 1967.

\bibitem{glickman2012example}
M.~E. Glickman, ``Example of the glicko-2 system,'' \emph{Boston University}, vol.~28, 2012.

\bibitem{herbrich2006trueskill}
R.~Herbrich, T.~Minka, and T.~Graepel, ``Trueskill™: a bayesian skill rating system,'' \emph{Advances in neural information processing systems}, vol.~19, 2006.

\bibitem{clubelo}
\BIBentryALTinterwordspacing
L.~Schiefler. Club elo. [Online]. Available: \url{http://www.clubelo.com}
\BIBentrySTDinterwordspacing

\bibitem{uefa}
\BIBentryALTinterwordspacing
Uefa rating. [Online]. Available: \url{https://www.uefa.com/nationalassociations/uefarankings/country}
\BIBentrySTDinterwordspacing

\bibitem{vevcek2014chess}
N.~Ve{\v{c}}ek, M.~Mernik, and M.~{\v{C}}repin{\v{s}}ek, ``A chess rating system for evolutionary algorithms: A new method for the comparison and ranking of evolutionary algorithms,'' \emph{Information Sciences}, vol. 277, pp. 656--679, 2014.

\bibitem{dehpanah2021evaluation}
A.~Dehpanah, M.~F. Ghori, J.~Gemmell, and B.~Mobasher, ``The evaluation of rating systems in online free-for-all games,'' in \emph{Advances in Data Science and Information Engineering: Proceedings from ICDATA 2020 and IKE 2020}.\hskip 1em plus 0.5em minus 0.4em\relax Springer, 2021, pp. 131--151.

\bibitem{glickman2018comparison}
M.~E. Glickman, J.~Hennessy, and A.~Bent, ``A comparison of rating systems for competitive women's beach volleyball,'' \emph{Statistica Applicata-Italian Journal of Applied Statistics}, no.~2, pp. 233--254, 2018.

\bibitem{ebtekar2021elo}
A.~Ebtekar and P.~Liu, ``An elo-like system for massive multiplayer competitions,'' \emph{arXiv preprint arXiv:2101.00400}, 2021.

\bibitem{park2021online}
J.~Park, ``Online estimation of student ability and item difficulty with glicko-2 rating system on stratified data.'' \emph{International Educational Data Mining Society}, 2021.

\bibitem{olsson2018skill}
C.~Olsson, S.~Bhupatiraju, T.~Brown, A.~Odena, and I.~Goodfellow, ``Skill rating for generative models,'' \emph{arXiv preprint arXiv:1808.04888}, 2018.

\bibitem{cardoso2020decoding}
L.~F. Cardoso, V.~C. Santos, R.~S.~K. Franc{\^e}s, R.~B. Prud{\^e}ncio, and R.~C. Alves, ``Decoding machine learning benchmarks,'' in \emph{Brazilian Conference on Intelligent Systems}.\hskip 1em plus 0.5em minus 0.4em\relax Springer, 2020, pp. 412--425.

\bibitem{dixon1997modelling}
M.~J. Dixon and S.~G. Coles, ``Modelling association football scores and inefficiencies in the football betting market,'' \emph{Journal of the Royal Statistical Society: Series C (Applied Statistics)}, vol.~46, no.~2, pp. 265--280, 1997.

\end{thebibliography}

% \clearpage\section*{Appendix}
\appendices
\section{Tables with ratings}

\clearpage

\begin{table}[h!]
\caption{Ratings of the best European clubs}
    % \centering
    \begin{tabular}{|l|l|l|l|}
    \hline
        \# & team & rating & league \\ \hline
        1 & Manchester City & 2237.7 & England. First \\ \hline
        2 & Bayern Munich & 2175.0 & Germany. First \\ \hline
        3 & FC Porto & 2135.9 & Portugal. First \\ \hline
        4 & Real Madrid & 2125.4 & Spain. First \\ \hline
        5 & Napoli & 2121.7 & Italy. First \\ \hline
        6 & Inter & 2116.4 & Italy. First \\ \hline
        7 & Liverpool & 2111.2 & England. First \\ \hline
        8 & Arsenal & 2101.2 & England. First \\ \hline
        9 & Manchester Utd & 2099.5 & England. First \\ \hline
        10 & RB Leipzig & 2096.4 & Germany. First \\ \hline
        11 & Benfica Portugal & 2093.1 & Portugal. First \\ \hline
        12 & Barcelona & 2078.6 & Spain. First \\ \hline
        13 & Salzburg & 2078.2 & Austria. First \\ \hline
        14 & Dortmund & 2077.2 & Germany. First \\ \hline
        15 & Paris SG & 2068.2 & France. First \\ \hline
        16 & AC Milan & 2047.5 & Italy. First \\ \hline
        17 & Newcastle & 2044.3 & England. First \\ \hline
        18 & Atl. Madrid & 2039.6 & Spain. First \\ \hline
        19 & Juventus & 2033.9 & Italy. First \\ \hline
        20 & Crvena zvezda & 2023.9 & Serbia. First \\ \hline
        21 & PSV & 2022.3 & Netherlands. First \\ \hline
        22 & Celtic & 2015.7 & Scotland. First \\ \hline
        23 & Sporting CP & 2015.4 & Portugal. First \\ \hline
        24 & Feyenoord & 2014.1 & Netherlands. First \\ \hline
        25 & Rangers Scotland & 2010.9 & Scotland. First \\ \hline
        26 & Lens & 2006.4 & France. First \\ \hline
        27 & Zenit & 2006.2 & Russia. First \\ \hline
        28 & Shakhtar Donetsk & 2004.7 & Ukraine. First \\ \hline
        29 & Young Boys & 2003.6 & Switzerland. First \\ \hline
        30 & Union Berlin & 2001.2 & Germany. First \\ \hline
        31 & Lazio & 2000.3 & Italy. First \\ \hline
        32 & Freiburg & 1999.8 & Germany. First \\ \hline
        33 & Real Sociedad & 1995.0 & Spain. First \\ \hline
        34 & Tottenham & 1991.5 & England. First \\ \hline
        35 & Ajax & 1991.2 & Netherlands. First \\ \hline
        36 & Brighton & 1990.1 & England. First \\ \hline
        37 & AS Roma & 1988.2 & Italy. First \\ \hline
        38 & Rakow & 1982.7 & Poland. First \\ \hline
        39 & Fenerbahce & 1978.0 & Turkey. First \\ \hline
        40 & Chelsea & 1976.6 & England. First \\ \hline
        41 & Atalanta & 1973.5 & Italy. First \\ \hline
        42 & Bayer Leverkusen & 1973.0 & Germany. First \\ \hline
        43 & Eintracht Frankfurt & 1968.0 & Germany. First \\ \hline
        44 & Braga & 1967.6 & Portugal. First \\ \hline
        45 & Aston Villa & 1967.5 & England. First \\ \hline
        46 & Royale Union SG & 1963.4 & Belgium. First \\ \hline
        47 & Marseille & 1962.1 & France. First \\ \hline
        48 & Brentford & 1958.7 & England. First \\ \hline
        49 & Villarreal & 1958.4 & Spain. First \\ \hline
        50 & Galatasaray & 1955.9 & Turkey. First \\ \hline
    \end{tabular}
\end{table}

\newpage

\begin{table}[h!]
\caption{Ratings of the best South American clubs}
    % \centering
    \begin{tabular}{|l|l|l|l|}
    \hline
        \# & team & rating & league \\ \hline
        1 & Palmeiras & 1961.0 & Brazil. First \\ \hline
        2 & Flamengo RJ & 1883.1 & Brazil. First \\ \hline
        3 & Atletico-MG & 1849.4 & Brazil. First \\ \hline
        4 & River Plate Argentina & 1842.7 & Argentina. First \\ \hline
        5 & Athletico-PR & 1826.8 & Brazil. First \\ \hline
        6 & Ind. del Valle & 1823.2 & Ecuador. First \\ \hline
        7 & Fluminense & 1821.5 & Brazil. First \\ \hline
        8 & Libertad Asuncion & 1815.7 & Paraguay. First \\ \hline
        9 & Internacional & 1810.3 & Brazil. First \\ \hline
        10 & Sao Paulo & 1809.9 & Brazil. First \\ \hline
        11 & Cerro Porteno & 1793.6 & Paraguay. First \\ \hline
        12 & Defensa y Justicia & 1787.2 & Argentina. First \\ \hline
        13 & Fortaleza Brazil & 1782.6 & Brazil. First \\ \hline
        14 & Botafogo RJ & 1780.7 & Brazil. First \\ \hline
        15 & Talleres Cordoba & 1775.5 & Argentina. First \\ \hline
        16 & Nacional Uruguay & 1772.8 & Uruguay. First \\ \hline
        17 & LDU Quito & 1749.1 & Ecuador. First \\ \hline
        18 & Boca Juniors & 1748.4 & Argentina. First \\ \hline
        19 & Olimpia Asuncion & 1747.6 & Paraguay. First \\ \hline
        20 & Estudiantes L.P. & 1745.8 & Argentina. First \\ \hline
        21 & Bragantino & 1742.5 & Brazil. First \\ \hline
        22 & Gremio & 1740.8 & Brazil. First \\ \hline
        23 & Corinthians & 1737.5 & Brazil. First \\ \hline
        24 & San Lorenzo Argentina & 1736.6 & Argentina. First \\ \hline
        25 & Millonarios & 1732.7 & Colombia. First \\ \hline
        26 & Newells Old Boys & 1728.4 & Argentina. First \\ \hline
        27 & Racing Club & 1723.9 & Argentina. First \\ \hline
        28 & Atl. Nacional & 1718.7 & Colombia. First \\ \hline
        29 & Belgrano & 1717.7 & Argentina. First \\ \hline
        30 & Argentinos Jrs & 1712.4 & Argentina. First \\ \hline
        31 & Rosario Central & 1704.5 & Argentina. First \\ \hline
        32 & America MG & 1704.4 & Brazil. First \\ \hline
        33 & Cruzeiro & 1701.2 & Brazil. First \\ \hline
        34 & Atletico GO & 1699.1 & Brazil. Second \\ \hline
        35 & Sport Recife & 1697.7 & Brazil. Second \\ \hline
        36 & Barcelona SC & 1694.3 & Ecuador. First \\ \hline
        37 & Ceara & 1688.8 & Brazil. Second \\ \hline
        38 & Penarol & 1688.2 & Uruguay. First \\ \hline
        39 & Godoy Cruz & 1685.2 & Argentina. First \\ \hline
        40 & Goias & 1681.7 & Brazil. First \\ \hline
        41 & Nacional Asuncion & 1681.1 & Paraguay. First \\ \hline
        42 & Guarani Paraguay & 1681.0 & Paraguay. First \\ \hline
        43 & Cuiaba & 1680.3 & Brazil. First \\ \hline
        44 & Lanus & 1680.2 & Argentina. First \\ \hline
        45 & Tigre & 1677.0 & Argentina. First \\ \hline
        46 & Sportivo Trinidense & 1676.8 & Paraguay. First \\\hline
        47 & Santos & 1674.6 & Brazil. First \\ \hline
        48 & U. Catolica Ecuador & 1667.5 & Ecuador. First \\ \hline
        49 & Aucas & 1667.0 & Ecuador. First \\ \hline
        50 & Bahia & 1664.5 & Brazil. First \\ \hline
    \end{tabular}
\end{table}

\begin{table}[!ht]
\caption{Average European league ratings of top 5 teams}
    \centering
    \begin{tabular}{|l|l|l|l|l|l|}
    \hline
        \# & league & rating & \# & league & rating \\ \hline
        1 & England. First & 2118.8 & 50 & Lithuania. First & 1661.7 \\ \hline
        2 & Germany. First & 2069.9 & 51 & North Macedonia. First & 1651.8 \\ \hline
        3 & Italy. First & 2064.0 & 52 & Portugal. Second & 1646.9 \\ \hline
        4 & Spain. First & 2039.4 & 53 & Kosovo. First & 1642.1 \\ \hline
        5 & Portugal. First & 2000.7 & 54 & Luxembourg. First & 1640.6 \\ \hline
        6 & France. First & 1981.4 & 55 & Estonia. First & 1637.8 \\ \hline
        7 & Netherlands. First & 1965.8 & 56 & Moldova. First & 1636.9 \\ \hline
        8 & Belgium. First & 1923.3 & 57 & Malta. First & 1622.7 \\ \hline
        9 & Turkey. First & 1911.9 & 58 & Austria. Second & 1619.6 \\ \hline
        10 & Russia. First & 1906.5 & 59 & Denmark. Second & 1616.9 \\ \hline
        11 & Austria. First & 1899.3 & 60 & Faroe Islands. First & 1615.1 \\ \hline
        12 & Switzerland. First & 1893.4 & 61 & Albania. First & 1614.2 \\ \hline
        13 & Poland. First & 1890.7 & 62 & Gibraltar. First & 1612.3 \\ \hline
        14 & Sweden. First & 1883.9 & 63 & Greece. Second & 1611.2 \\ \hline
        15 & Ukraine. First & 1878.1 & 64 & Netherlands. Second & 1602.1 \\ \hline
        16 & Greece. First & 1868.8 & 65 & Sweden. Second & 1587.9 \\ \hline
        17 & Denmark. First & 1852.3 & 66 & Belgium. Second & 1581.6 \\ \hline
        18 & Czech Republic. First & 1832.4 & 67 & Romania. Second & 1577.1 \\ \hline
        19 & Scotland. First & 1826.2 & 68 & Montenegro. First & 1562.7 \\ \hline
        20 & Israel. First & 1817.1 & 69 & Hungary. Second & 1557.3 \\ \hline
        21 & Serbia. First & 1816.6 & 70 & Wales. First & 1556.8 \\ \hline
        22 & Bulgaria. First & 1809.7 & 71 & Israel. Second & 1553.7 \\ \hline
        23 & Norway. First & 1808.1 & 72 & Scotland. Second & 1532.3 \\ \hline
        24 & Romania. First & 1807.6 & 73 & Croatia. Second & 1525.0 \\ \hline
        25 & Croatia. First & 1807.2 & 74 & Ukraine. Second & 1519.7 \\ \hline
        26 & Cyprus. First & 1795.4 & 75 & Slovenia. Second & 1517.7 \\ \hline
        27 & Germany. Second & 1785.3 & 76 & Norway. Second & 1516.7 \\ \hline
        28 & Latvia. First & 1767.9 & 77 & Slovakia. Second & 1490.9 \\ \hline
        29 & Italy. Second & 1767.6 & 78 & Kazakhstan. Second & 1477.3 \\ \hline
        30 & England. Second & 1766.0 & 79 & Serbia. Second & 1473.3 \\ \hline
        31 & Spain. Second & 1763.3 & 80 & North Macedonia. Second & 1466.5 \\ \hline
        32 & Slovenia. First & 1749.0 & 81 & Bulgaria. Second & 1460.6 \\ \hline
        33 & Slovakia. First & 1748.5 & 82 & Georgia. Second & 1456.8 \\ \hline
        34 & Belarus. First & 1735.6 & 83 & Bosnia And Herzegovina. Second & 1456.6 \\ \hline
        35 & Armenia. First & 1732.3 & 84 & Czech Republic. Second & 1453.2 \\ \hline
        36 & Hungary. First & 1727.1 & 85 & Andorra. First & 1443.9 \\ \hline
        37 & Georgia. First & 1726.1 & 86 & Armenia. Second & 1443.3 \\ \hline
        38 & Bosnia And Herzegovina. First & 1724.8 & 87 & Cyprus. Second & 1436.2 \\ \hline
        39 & Finland. First & 1721.9 & 88 & Finland. Second & 1430.4 \\ \hline
        40 & France. Second & 1721.9 & 89 & Ireland. Second & 1427.2 \\ \hline
        41 & Northern Ireland. First & 1706.5 & 90 & San Marino. First & 1412.1 \\ \hline
        42 & Azerbaijan. First & 1704.0 & 91 & Belarus. Second & 1409.0 \\ \hline
        43 & Switzerland. Second & 1700.7 & 92 & Albania. Second & 1374.1 \\ \hline
        44 & Ireland. First & 1698.4 & 93 & Latvia. Second & 1365.0 \\ \hline
        45 & Turkey. Second & 1689.6 & 94 & Iceland. Second & 1362.8 \\ \hline
        46 & Russia. Second & 1684.8 & 95 & Malta. Second & 1327.4 \\ \hline
        47 & Kazakhstan. First & 1683.2 & 96 & Lithuania. Second & 1316.7 \\ \hline
        48 & Poland. Second & 1671.4 & 97 & Estonia. Second & 1304.2 \\ \hline
        49 & Iceland. First & 1669.8 & 98 & Northern Ireland. Second & 1276.8 \\ \hline
    \end{tabular}
\end{table}

\begin{table}[!ht]
\caption{Average South American league ratings of top 5 teams}
    \centering
    \begin{tabular}{|l|l|l|}
    \hline
        \# & league & rating \\ \hline
        1 & Brazil. First & 1868.4 \\ \hline
        2 & Argentina. First & 1779.9 \\ \hline
        3 & Paraguay. First & 1743.8 \\ \hline
        4 & Ecuador. First & 1720.2 \\ \hline
        5 & Brazil. Second & 1676.0 \\ \hline
        6 & Uruguay. First & 1673.1 \\ \hline
        7 & Colombia. First & 1665.6 \\ \hline
        8 & Chile. First & 1605.5 \\ \hline
        9 & Argentina. Second & 1604.4 \\ \hline
        10 & Peru. First & 1592.5 \\ \hline
        11 & Venezuela. First & 1557.8 \\ \hline
        12 & Bolivia. First & 1556.2 \\ \hline
    \end{tabular}
\end{table}

\end{document}